\documentclass[aps,prb,twocolumn,groupedaddress,pdflatex,floatfix]{revtex4-2}
\pdfoutput=1

\usepackage{amsmath}
\usepackage{comment}
\usepackage{graphicx}
\usepackage[colorlinks=true,allcolors=blue]{hyperref}
\usepackage{xcolor}
\usepackage[caption=false]{subfig}
\newcommand{\phantomsubfloat}[1]{
    {
        \captionsetup[subfigure]{labelformat=empty}
        \subfloat[][]{#1}
    }%
}

\begin{document}


\title{Optimizing the transport of Majorana zero modes in one-dimensional topological superconductors}


\author{Bill P. Truong}
	\email[]{bill.truong@mail.mcgill.ca}

\author{Kartiek Agarwal}

\author{T. Pereg-Barnea}

\affiliation{Department of Physics and the Centre for the Physics of Materials, McGill University, Montr\'{e}al, Qu\'{e}bec H3A 2T8, Canada}


\date{\today}

\begin{abstract}
Topological quantum computing is based on the notion of braiding non-Abelian anyons, such as Majorana zero modes (MZMs), to perform gate operations. A crucial building block of these protocols is the adiabatic shuttling of MZMs through topological superconductors. Here, we consider the ``piano key'' approach, where MZMs are transported using local electric gates to tune sections (``keys'') of a wire between topologically trivial and nontrivial phases. We numerically simulate this transport on a single wire and calculate the diabatic error corresponding to exciting the system. We find that this error is typically reduced when transport is facilitated by using multiple keys as one may expect from modeling each piano key press as an effective Landau-Zener process. However, further increasing the number of keys increases errors; thus, there exists a nontrivial optimal number of keys that minimizes the diabatic error given a fixed total shuttle time. As we show, this optimal number of keys can be explained by modeling each key press as an effective Landau-Zener process while paying careful attention to power-law corrections that arise due to the nonanalytic behavior of the time-dependent modulation of the chemical potential at the beginning and end of each key press. 
\end{abstract}


\maketitle

\section{Introduction} \label{sec:introduction}
Majorana zero modes (MZMs) provide a promising platform for quantum computation due to their nonlocal character and non-Abelian exchange statistics \cite{Nayak2008, DasSarma2015}. Quantum information can be encoded nonlocally in a degenerate ground-state subspace formed by MZMs and is topologically protected against decoherence \cite{Kitaev2001, DasSarma2015}. The braiding of MZMs equates to a unitary operation within this subspace, which can be exploited for use as quantum logic gates \cite{Ivanov2001, Kitaev2003}. MZMs are predicted to emerge as edge states in $p$-wave superconductors \cite{Kitaev2001, Ivanov2001} and various experimental platforms for realizing $p$-wave superconductivity have been proposed \cite{Fu2008, Lutchyn2010, Oreg2010, Sau2010_1, Sau2010_2, Stanescu2011, Cook2011, Nadj-Perge2013}. Experimental signatures of MZMs have also been well-established, with a notable signature being a quantized electrical conductance at zero-bias voltage \cite{Law2009, Flensberg2010, Liu2012, DasSarma2012}. Over the past decade, considerable progress has been made in the experimental detection of MZMs in a host of settings including hybrid semiconductor-superconductor nanowires \cite{Mourik2012, Lee2012, Das2012, Deng2012, Finck2013, Churchill2013, Lee2014, Albrecht2016, Deng2016, Nichele2017, Gul2018, Deng2018, Vaitiekenas2020}, magnetic atomic chains \cite{Nadj-Perge2014, Jeon2017, Wang2021, Fan2021, Schneider2022}, and others \cite{Xu2015, Lv2017, Sun2016, Manna2020, Liu2020}. Although definitive evidence for MZMs remains elusive, recent experimental developments are encouraging and have galvanized theoretical efforts regarding their applications to quantum computation, with particular attention directed towards braiding. Various schemes for braiding MZMs have been proposed which involve circuits of superconducting wires \cite{Alicea2011, Sau2011, Halperin2012, Tutschku2020}, Josephson junctions \cite{vanHeck2012, Hyart2013, Hegde2020}, periodic driving \cite{Bauer2019, Martin2020, Min2022}, and others \cite{Li2016, Karzig2017, Malciu2018, Trif2019, Sanno2021}. 

The method by which MZMs are manipulated and the time scales over which these manipulations occur are important considerations for any braiding protocol. In particular, a number of protocols rely on MZMs being transported across superconducting wires. Transport which is performed too quickly compared to the time scale of the energy gap runs the risk of inducing quasiparticle excitations between the ground-state subspace and excited states. These diabatic transitions are a source of decoherence and are therefore destructive in view of topological quantum computation. Errors originating from such transitions have been examined in the specific case of transport \cite{Scheurer2013, Karzig2013, Karzig2015_2, Bauer2018, Conlon2019, Coopmans2021, Xu2022} and in the broader context of braiding \cite{Cheng2011, Karzig2015_1, Knapp2016, Hell2016, Rahmani2017, Sekania2017, Nag2019, Zhang2019}.

In this work, we study the diabatic error that is accrued as MZMs are transported over a fixed distance along a single, topological superconducting wire. We specifically focus on the ``piano key'' set-up where a wire is divided into electrically gated sections (``keys''). Within each key, the chemical potential may be tuned in order for the key to switch from a topologically trivial phase to a nontrivial phase and vice-versa. Since MZMs reside at the phase boundaries, appropriately tuning a key facilitates their transport. Our work is concerned with the use of multiple keys and serves as an extension of Ref. \cite{Bauer2018} which examines the diabatic error generated when a single key is used. We note that recent experiments have demonstrated more advanced electrical gating techniques \cite{Vaitiekenas2018, deMoor2018} which has led to an increased appeal in this transport method for theoretical study. Other methods have also been proposed where, for example, the chemical potential changes in a domain wall-like fashion \cite{Karzig2013, Karzig2015_2}.

While it is sufficient to use a single piano key to shuttle an MZM between two positions in a wire, it can nevertheless be useful to study how the diabatic error changes when this single key is separated into multiple keys which are tuned one at a time. We consider a simple protocol where an MZM is transported over a distance $R$ in a total time $\tau$ using $n$ keys where each key has the same size $R/n$ and the same tuning time $\tau/n$. The total diabatic error can be intuitively understood by viewing the tuning of each key as a Landau-Zener process \cite{Bauer2018}. As a key undergoes a topological phase transition, the energy gap between the ground state subspace and excited states achieves a minimum value $\delta \sim n/R$. Since Landau-Zener theory predicts that the total diabatic error scales as $\sim n e^{- \delta^2 \tau /v}$ for some model-dependent parameter $v$, it would appear that reducing the size of the keys (at least down to the MZM localization length) is always optimal. 

As we show, the above argument ignores important power-law corrections $\sim 1/(\delta \tau)^\alpha$ to the Landau-Zener result. These corrections originate due to the finite duration of the protocol and the nonanalytic behavior of the change in the chemical potential with time, $\mu(t)$, at the beginning and end of each piano key press. Somewhat counterintuitively, as the protocol time $\tau$ is increased, these corrections dominate the diabatic error as the usual exponentially small contribution becomes irrelevant. These power-law corrections can be suppressed by choosing a tuning function for $\mu(t)$ where the first $M$ derivatives vanish at the endpoints of each key press (leading to $\alpha = 2(M+1)$), though this is expected to be futile in the presence of noise~\cite{Knapp2016}.

In view of this, we consider piano keys for which the local chemical potential $\mu(t)$ is simply rescaled temporally to go from one extremal value to the other in time $\tau/n$. In this setting, there is a competition between the usual exponentially small contribution to the diabatic error expected from a Landau-Zener process, which favors increasing the number of keys, and the anomalous power-law contribution which favors decreasing the number of keys, and which dominates when larger protocol times are allowed. Thus, there exists a nontrivial number of keys $n^{*}$ which is optimal at reducing the diabatic error given either a fixed total protocol time or desired error tolerance. Interestingly, $n^{*} \rightarrow 1$ in the limit where the total protocol time is not a relevant constraint. 

This paper is organized as follows. In Sec. \ref{sec:setup}, we review the Kitaev chain as a model for a topological superconducting wire and describe the piano key protocol that we study for adiabatically transporting MZMs. Furthermore, we determine analytical expressions for the diabatic error by considering the dynamics of a two-level system undergoing Landau-Zener transitions. This analysis is undertaken for two chemical potential tuning functions which have different degrees of nonanalyticity at their endpoints, leading to different power-law corrections. In Sec. \ref{sec:results}, we demonstrate our numerical results for the diabatic error obtained from simulations in the case of multiple piano keys. We show that these results compare well to analytical predictions. Moreover, we show the emergence of an optimal number of piano keys, $n^{*}$, depending on the total protocol time or desired error tolerance. In Sec. \ref{sec:conclusion}, we conclude with a summary of our results.

\begin{figure}[t]
	\centering
  	\includegraphics[width=1.0\linewidth]{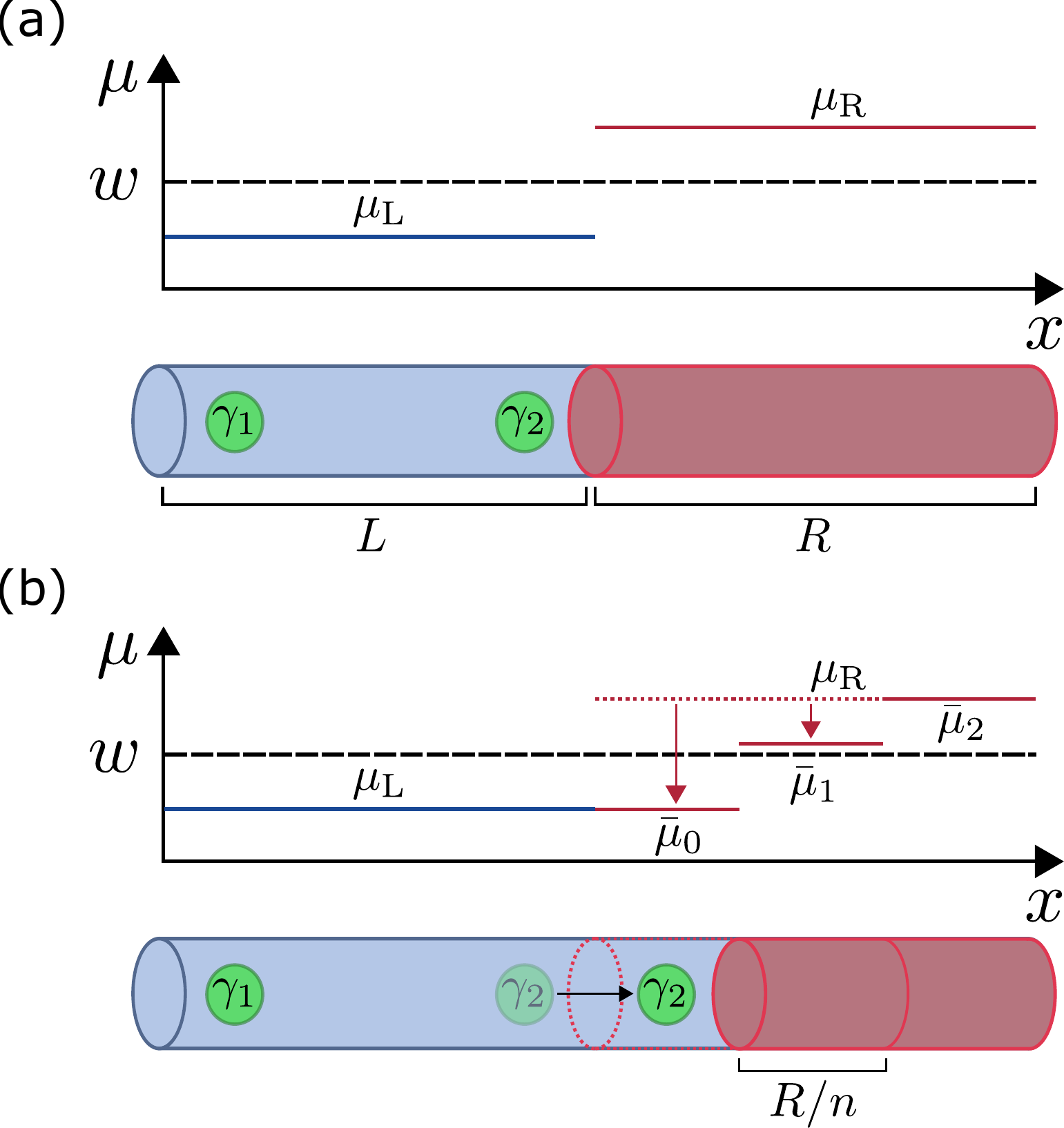}
  	\phantomsubfloat{\label{fig:pianokey_schematic_a}}
  	\phantomsubfloat{\label{fig:pianokey_schematic_b}}
  	\vspace{-2\baselineskip}
    \caption[]{Set-up for the transport of MZMs in a superconducting wire using the piano key approach. (a) The initial configuration of the wire which is divided into two sections. The left section is in the topologically nontrivial phase with length $L$ and chemical potential $\mu_{\text{L}}$ (solid blue line). The right section is in the trivial phase with length $R$ and chemical potential $\mu_{\text{R}}$ (solid red line). The critical value of the chemical potential which separates both phases is $\mu_{c} = w$ (dashed black line) for the Kitaev chain. The MZMs are represented as green circles and are labeled as $\gamma_{1}$ and $\gamma_{2}$. (b) The MZM $\gamma_{2}$ within the bulk of the wire is shuttled to the right by tuning the chemical potentials $\bar{\mu}_{m}(t)$ for $n$ keys each with length $R/n$. The chemical potentials are tuned from a value $\mu_{\text{R}}$ to $\mu_{\text{L}}$. Illustrated is the specific case of $n = 3$ with the first key having been completely pressed while the second key is in motion.}
  	\label{fig:pianokey_schematic}
\end{figure}

\section{Set-up and diabatic error} \label{sec:setup}
\subsection{Model for a superconducting wire} \label{sec:setup:subsec:model}

The model that we use for a one-dimensional, spinless $p$-wave superconductor is the Kitaev chain \cite{Kitaev2001}. In terms of electron operators, the Hamiltonian for a Kitaev chain with $N$ electronic sites is 
\begin{align}
	H_{\text{kit}} &= - \sum_{j=1}^{N} \mu_{j} c_{j}^{\dagger} c_{j} - \frac{w}{2} \sum_{j=1}^{N-1} (c_{j}^{\dagger} c_{j+1} + \text{H.c.})
	\label{eq:ham_kitaev_realspace}
	\\
	&- \frac{\Delta_{\text{SC}}}{2} \sum_{j=1}^{N-1} (c_{j} c_{j+1} + \text{H.c.}),
	\nonumber
\end{align}
where $\mu_{j}$ is the site-dependent chemical potential, $w > 0$ is the nearest-neighbour hopping amplitude, and $\Delta_{\text{SC}} > 0$ is the superconducting pairing amplitude. Equation (\ref{eq:ham_kitaev_realspace}) may also be cast into a form which uses Majorana operators; see Appendix \ref{app:majorana} for details. For a chain with homogeneous chemical potential $\mu_{j} = \mu$, MZMs appear when the chain is in the topologically nontrivial phase, which occurs when $|\mu| < w$. The MZMs are localized at the edges, exponentially decay into the bulk, and display an energy splitting which is exponentially suppressed by their separation distance. Conversely, the topologically trivial phase corresponds to $|\mu| > w$, which does not feature MZMs. 

Generally, a chain with inhomogeneous parameters may consist of sections which are in different topological phases. At the boundaries of these different phases are where MZMs are pinned. We consider a chain which is divided into two sections: the left section, which has length $L$, is placed in the nontrivial phase and the right section, which has length $R$, can be interpolated between both phases. Initially, the right section is placed in the trivial phase, which results in the appearance of one MZM at the far left edge of the chain (denoted by $\gamma_{1}$) and another in the bulk (denoted by $\gamma_{2}$). The initial setup of the chain is illustrated in Fig. \ref{fig:pianokey_schematic_a}. Our work focuses specifically on the transport of the MZM $\gamma_{2}$ to the far right edge by controlling the phase of the right section via the chemical potential. 

Let us consider quantitatively how the chain parameters change with position and time. We set the parameters $w$ and $\Delta_{\text{SC}}$ to be time-independent. For the left section to be fixed in the nontrivial phase, its chemical potential is chosen to be $\mu_{j} = \mu_{\text{L}}$ for $j \leq L$, where $|\mu_{\text{L}}| < w$. For the right section to be initially in the trivial phase, its chemical potential is chosen to be $\mu_{j} = \mu_{\text{R}}$ for $j = L < 0 \leq N$, where $|\mu_{\text{R}}| > w$. As previously discussed, the different sections of the chain can be thought of as ``piano keys'' and moving $\gamma_{2}$ requires keys to be ``pressed.'' We first review the case where the right section is treated as a single key and adopt the scheme detailed in Ref. \cite{Bauer2018} to tune the chemical potential. The chemical potential changes with time uniformly across the entire key: $\mu_{j} = \mu(t)$ for $j = L < 0 \leq N$. Here, $\mu(t)$ is adjusted from $\mu_{\text{R}}$ to $\mu_{\text{L}}$ over a total time $\tau$ and takes the form
\begin{equation}
	\mu(t) = [1 - g(t/\tau)] \mu_{\text{L}} + g(t/\tau) \mu_{\text{R}},
	\label{eq:mu_time}
\end{equation}
with
\begin{equation}
	g(t/\tau) = 
	\begin{cases}
		0 & t/\tau \leq 0 
		\\
		f(t/\tau) & 0 < t/\tau < 1
		\\
		1 & t/\tau \geq 1
	\end{cases},
	\label{eq:func_g}
\end{equation}
where $f(t/\tau)$ is a tuning function such that $f(0) = 0$ and $f(1) = 1$. In later sections, we consider both a linear tuning function $f(t/\tau) = t/\tau$ as well as a ``smooth'' tuning function $f(t/\tau) = \sin^2(\pi t/2 \tau)$. 

Our work is predominantly concerned with the case of transport using multiple piano keys, as illustrated in Fig. \ref{fig:pianokey_schematic_b}. We consider now the division of the right section of the chain into $n$ keys each with equal length $R/n$. The MZM $\gamma_{2}$ may be shuttled to the far right edge by successively pressing each adjacent key beginning with the key located at the initial phase boundary. The amount of time allocated to pressing each key is given by $\tau/n$. We remark that as soon as one key completes a press, the next key is set into motion instantly. The scheme used to change the chemical potential for a single key, as described in Eqs. (\ref{eq:mu_time}) and (\ref{eq:func_g}), is adapted for each key in this case. Let the chemical potential of a key be denoted by $\bar{\mu}_{m}(t)$ where $m = 0, 1, 2, ..., n - 1$ labels a key's pressing order. The chemical potential of the $m$th key can be formulated as a rescaling and shifting of time applied to Eq. (\ref{eq:mu_time}):
\begin{equation}
	\bar{\mu}_{m}(t) = \mu(n t - m \tau).
	\label{eq:chempot_multiple}
\end{equation}
Both Eqs. (\ref{eq:func_g}) and (\ref{eq:chempot_multiple}) ensure that the chemical potential of the $m$th key is tuned only when $t \in [m\tau/n, (m+1)\tau/n]$. 

We briefly comment on the use of electric gates to tune each key in experiment. First, we note that the size of a key is controlled by the size of its corresponding gate. In practice, electric gates can be manufactured with a size on the scale of $\sim 10$ nm using methods such as electron beam lithography, see Refs. \cite{Groves2014} and \cite{Chen2015} for details. In our simulations, we consider a superconducting wire with a length of $\sim 1$-$2$ $\mu$m and study MZM transport using a maximum of $n = 15$ keys. This means that the smallest key that we consider has a size of $\sim 70$ nm which is within the realm of experimental feasibility. 


\subsection{Diabatic error of single piano key}\label{sec:model:subsec:single_pk}

As previously mentioned, the diabatic error is defined to be the transition probability between the ground state and the excited states. A general expression for the diabatic error may be given as
\begin{equation}
	\mathcal{P} = 1 - | \langle \Omega_{f} | U(\tau) | \Omega_{i} \rangle |^2,
	\label{eq:diaberr_gen}
\end{equation}
where $| \Omega_{i} \rangle$ is the initial ground state, $| \Omega_{f} \rangle$ is the instantaneous ground state of the final Hamiltonian, and $U(\tau)$ is a time evolution unitary. Within the context of this work, $U(\tau)$ encapsulates the details of transporting the MZM $\gamma_{2}$ over a total time $\tau$; see Appendix \ref{app:time} for details.

We review an analytical calculation of the diabatic error that is accumulated for a single piano key. As demonstrated in Ref. \cite{Bauer2018}, much of the underlying physics of the diabatic error can be captured by a simple Landau-Zener model. This calculation hinges on two simplifying assumptions. First, it is assumed that most of the contributions to the diabatic error originate from transitions between the ground state and the first excited state. Second, it is assumed that these transitions are most probable when the key is near the critical point that separates the two topologically distinct phases. 

In reference to the first assumption, let us define the ground state and first excited state. We consider the BCS ground state $| \Omega \rangle$ such that $d_{l} | \Omega \rangle = 0$, where $d_{l}$, $l = 0, 1, 2, ..., N-1$ are fermionic operators for Bogoliubov quasiparticles. The energy associated with creating a quasiparticle is $\epsilon_{l} > 0$ where $\epsilon_{0} < \epsilon_{1} < ... < \epsilon_{N-1}$. Since $H_{\text{kit}}$ conserves parity, we consider the first excited state that is within the same parity sector as the ground state.
This first excited state must be occupied by two quasiparticles, one with energy $\epsilon_{0}$ and another with energy $\epsilon_{1}$:
\begin{equation}
	| X \rangle = d_{0}^{\dagger} d_{1}^{\dagger} | \Omega \rangle.
	\label{eq:excitedstate}
\end{equation}
The operator $d_{0}$ is unique in that it contains both MZMs, namely $d_{0} = (1/2)(\gamma_{1} + i\gamma_{2})$. The energy of the MZMs is exponentially suppressed by their separation distance $\Delta x$ such that $\epsilon_{0} \sim e^{-\Delta x/\xi}$ where $\xi$ is the superconducting coherence length. When this energy splitting vanishes, which occurs in the large separation limit $\Delta x \gg \xi$ or when model parameters are fine-tuned, the opposite parity state $d_{0}^{\dagger} | \Omega \rangle$ becomes degenerate with $| \Omega \rangle$. In these limits, the set-up of the states presented can be equally applied to this opposite parity ground state. In either case, since $\epsilon_{0} \ll \epsilon_{1}$, the energy gap between the ground state and first excited state of the same parity can be well-approximated by $\epsilon_{1}$.

Invoking the second assumption, we consider the dynamics of the relevant eigenmodes when the piano key is in the vicinity of criticality. We briefly mention the behavior of the MZMs in this regime. As detailed in Ref. \cite{Bauer2018}, $\gamma_{1}$ on the far left edge is unaffected when criticality is achieved while $\gamma_{2}$  delocalizes across the region occupied by the key. When the key moves further into the nontrivial phase, $\gamma_{2}$ localizes on the far right edge. However, the bulk eigenmode represented by  $d_{1}$ also becomes localized within the piano key region in this critical regime. Its energy $\epsilon_{1}$ can be estimated by considering the key as its own separate chain with length $R$. If $R$ is sufficiently large, then this energy can be approximated by the expression for the bulk spectrum:
\begin{equation}
	\epsilon(k) = \sqrt{[\mu(t) + w \cos k]^2 + \Delta_{\text{SC}}^2 \sin^2 k},
	\label{eq:kit_bulkenergy}
\end{equation}
where $k$ is the momentum. Specializing to $\mu(t) > 0$, criticality occurs when $\mu(t) = w$ with the bulk gap closing at momentum $k = \pi$. Knowing that the momentum resolution is $\pi/R$ due to the finite size of the key, the lowest energy bulk mode is expected to have momentum $k = \pi + \pi/R$. The energy of this bulk mode, $\epsilon(\pi + \pi/R)$, is precisely what we seek. Assuming that $\pi/R \ll 1$, we expand the terms $\cos(\pi + \pi/R) \approx -1$ and $\sin(\pi + \pi/R) \approx - \pi/R$ in Eq. (\ref{eq:kit_bulkenergy}), and subsequently obtain an estimate for $\epsilon_{1}$:
\begin{equation}
	\epsilon_{1} \approx \sqrt{[\mu(t) - w]^2 + \left(\frac{\pi \Delta_{\text{SC}}}{R}\right)^2}.
	\label{eq:e1_bulkenergy}
\end{equation}

The picture that we have so far is that of a two-level system with an energy gap given by Eq. (\ref{eq:e1_bulkenergy}). An effective Hamiltonian describing this low energy subspace is given as
\begin{equation}
	H_{\text{eff}} = \frac{1}{2}
	\begin{pmatrix}
		\mu(t) - w & \dfrac{\pi \Delta_{\text{SC}}}{R} 
		\\
		\dfrac{\pi \Delta_{\text{SC}} }{R}  & -\mu(t) + w
	\end{pmatrix}.
	\label{eq:ham_twolevel}
\end{equation}
The basis of this effective Hamiltonian consists of the ground state $| \Omega \rangle$ and the first excited state $| X \rangle$ of the Kitaev Hamiltonian in the case where the energy gap vanishes at criticality, which occurs when either $\Delta_{\text{SC}} = 0$ or $R \rightarrow \infty$. As will be discussed, relevant to the calculation of the diabatic error are the instantaneous eigenstates of Eq. (\ref{eq:ham_twolevel}) which represent the ground state and first excited state in the case of a finite energy gap of $\pi \Delta_{\text{SC}}/R$ at criticality.

\begin{figure*}[t]
    \centering
  	\includegraphics[width=1.0\linewidth]{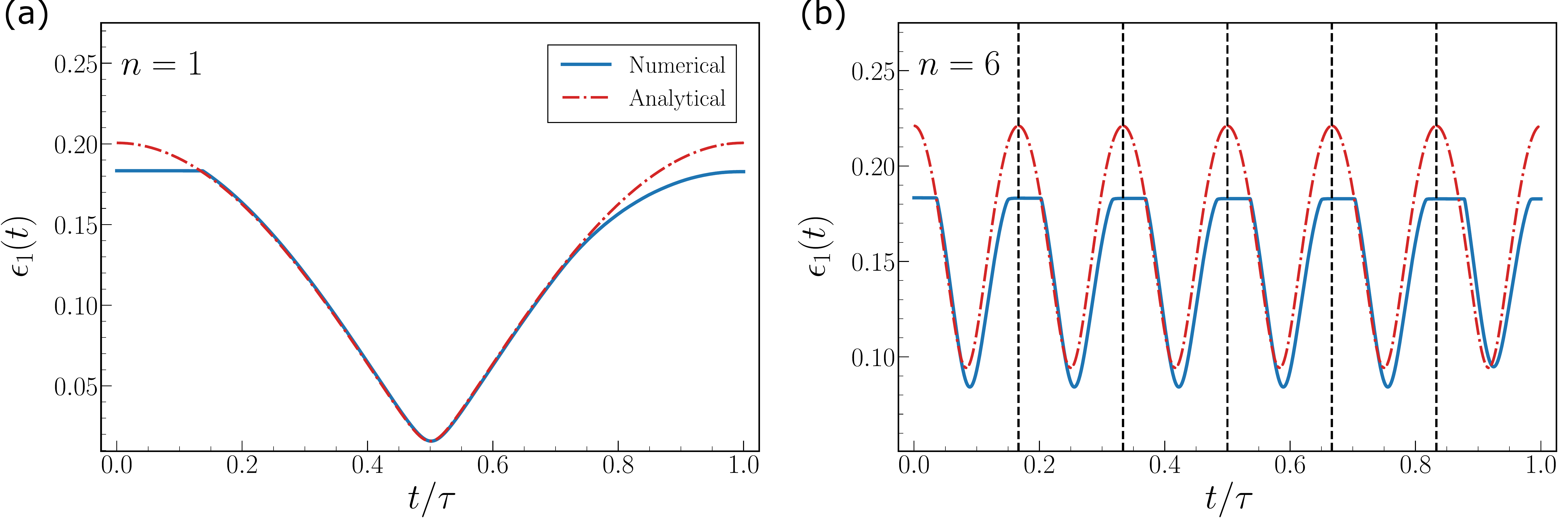}
  	\phantomsubfloat{\label{fig:energies_a}}
  	\phantomsubfloat{\label{fig:energies_b}}
  	\vspace{-2\baselineskip}
    \caption[]{Energy $\epsilon$ of the first excited state as the MZM $\gamma_{2}$ is transported using (a) a single piano key and (b) six piano keys. The numerical results (solid blue curves) were obtained using exact diagonalization for a Kitaev chain with $N = 240$ sites where the left and right sections are equal, $L = R = 120$ sites. A smooth tuning function $f(t/\tau) = \sin^2 (\pi t/2 \tau)$ was used to adjust the chemical potential in each key. The analytical expression for the energy from Eq. (\ref{eq:e1_bulkenergy}) is shown (red dot-dashed curves). For the single key protocol in panel (a), the analytical expression matches well with the numerical result in the critical regime. In the case of multiple keys in panel (b), the analytical expression overestimates the minimum gap for the first and intermediate keys, however it captures well the minimum gap of the final key. In numerical calculations, we use parameters $w = 3.0$ meV $\Delta_{\text{SC}} = 0.6$ meV, and $\delta \mu = 0.2$ meV.}
  	\label{fig:energies}
\end{figure*}

We now specialize to the case where the chemical potential changes symmetrically around the critical point, namely $\mu_{\text{L}} = w - \delta \mu$ and $\mu_{\text{R}} = w + \delta \mu$ for $\delta \mu > 0$. The effective Hamiltonian becomes
\begin{equation}
	H_{\text{eff}} = \frac{1}{2}
	\begin{pmatrix}
		\delta \mu [1 - 2 g(t/\tau)] & \dfrac{\pi \Delta_{\text{SC}}}{R} 
		\\
		\dfrac{\pi \Delta_{\text{SC}}}{R}  & - \delta \mu [1 - 2 g(t/\tau)]
	\end{pmatrix},
	\label{eq:ham_twolevel_symmetric}
\end{equation}
where we have explicitly used Eq. (\ref{eq:mu_time}). In Fig. \ref{fig:energies_a}, we compare the energy gap to the energy of the first excited state in the Kitaev chain as a single key is tuned using the established chemical potential configuration.

We first consider the case where the tuning function changes linearly with time: $f(t/ \tau) = t / \tau$. It is convenient to shift the time axis as $t \rightarrow t + \tau/2$ which results in criticality occurring at $t = 0$. The effective Hamiltonian reads
\begin{equation}
	H_{\text{eff,lin}} =
	\begin{pmatrix}
		- \dfrac{\delta \mu }{\tau} t & \dfrac{\pi \Delta_{\text{SC}}}{2 R} 
		\\[8.5pt]
		\dfrac{\pi \Delta_{\text{SC}}}{2 R}  & \dfrac{\delta \mu }{\tau}t
	\end{pmatrix},
	\label{eq:ham_twolevel_symmetric_linear}
\end{equation}
where $t \in [-\tau/2, \tau/2]$. The calculation of the diabatic error, as stated in Eq. (\ref{eq:diaberr_gen}), requires knowledge of the time evolution of the system. Here, the time evolution operator may be obtained by solving the Schr\"{o}dinger equation in the basis of the eigenstates of Eq. (\ref{eq:ham_twolevel_symmetric_linear}) with the initial condition $| \psi(0) \rangle \equiv | \Omega_{i} \rangle$. The exact expressions for the matrix elements of this time evolution operator have been calculated in Refs. \cite{Vitanov1996} and \cite{Vitanov1999} and take the form of parabolic cylinder functions. Using the results therein, we obtain an approximate expression for the diabatic error:
\begin{equation}
	P_{\text{lin}}(\tau) \sim e^{- \pi \Delta \tau / 2 r} + \left( \frac{r}{\tau \Delta} \right)^2 \frac{1}{(1 + r^2)^3}, 
	\label{eq:transProb_linear}
\end{equation}
where
\begin{equation}
	\Delta = \frac{\pi \Delta_{\text{SC}}}{2 R},~~r = \frac{\delta \mu}{2\Delta}.
	\label{eq:shorthand_linear}
\end{equation}
We note that the minimum gap is given by $2 \Delta$ while $r$ is roughly the ratio between the maximum and minimum gaps if $\Delta \ll \delta \mu$. The expression in Eq. (\ref{eq:transProb_linear}) is valid for $\tau \gg 4 \delta \mu / (\delta \mu ^2 + \Delta^2)$ which comes from the asymptotic expansion of the parabolic cylinder functions for simultaneously large argument and large parameter.

The diabatic error in this effective two-level system set-up features oscillations which are not captured by Eq. (\ref{eq:transProb_linear}). These oscillations originate from accidental closed orbits in the Bloch sphere which may reduce the diabatic error for fine-tuned parameters. We omit these oscillations since they are expected to be largely suppressed in the Kitaev chain simulations. This suppression stems from the fact that excitations from the ground state are more likely to leak into the continuum of excited states than to return back to the ground state itself. This is observed in our numerical results for the diabatic error, which will be discussed in Sec. \ref{sec:results}. Nonetheless, we have verified numerically that Eq. (\ref{eq:transProb_linear}) describes well the overall profile of the diabatic error in the case of the two-level system.

The exponentially decaying first term in Eq. (\ref{eq:transProb_linear}) is the familiar Landau-Zener formula and originates from the fact that the system contains a parabolic avoided level crossing at criticality. However, the correction term, which goes as a power-law in $\tau$, is directly related to how smoothly the energy levels change at the temporal endpoints of the tuning. The competition between both types of behavior crucially depends on the size of the minimum gap relative to the maximum gap as characterized by $r$: As the minimum gap grows, the power-law behavior begins to dominate at shorter $\tau$. 

The specific power-law which manifests as $\tau$ grows is highly dependent on the nonanalyticity of the tuning function at the temporal endpoints \cite{Garrido1962}. In particular, if the first $M$ time-derivatives of $f(t/\tau)$ vanish at the endpoints, then the diabatic error is expected to go as $\tau^{-2(M+1)}$. For the case of linear tuning, the first time-derivative is nonzero at these endpoints which leads to $M = 0$ and so $\tau^{-2}$ as expected. 

We also consider the smooth tuning function $f(t/\tau) = \sin^2 (\pi t/2 \tau)$. We shift the time axis in the same manner as for the linear tuning function and from Eq. (\ref{eq:ham_twolevel_symmetric}), obtain the following effective Hamiltonian:
\begin{equation}
	H_{\text{eff,sin}} =
	\begin{pmatrix}
		-\dfrac{\delta \mu}{2} \sin \left( \dfrac{\pi t}{\tau} \right) & \dfrac{\pi \Delta_{\text{SC}}}{2 R} 
		\\[10pt]
		\dfrac{\pi \Delta_{\text{SC}}}{2 R}  & \dfrac{\delta \mu}{2} \sin \left( \dfrac{\pi t}{\tau} \right)
	\end{pmatrix}.
	\label{eq:ham_twolevel_symmetric_sin}
\end{equation}
Using seminumerical methods augmented with the general form of the expressions found in Ref. \cite{Vitanov1999}, we determine a suitable approximation for the diabatic error:
\begin{equation}
	P_{\text{sin}}(\tau) \sim e^{- \Delta \tau / r} + \frac{6 r^2}{(\tau \Delta)^4} \frac{1}{(1 + r^2)^4}.
	\label{eq:transProb_sin}
\end{equation}
As with the linear case, the first term of Eq. (\ref{eq:transProb_sin}) corresponds to the Landau-Zener formula while the second term is a power-law correction. The $\tau^{-4}$ dependence is consistent with expectations from general theory. Here, the first time-derivative of $f(t/\tau)$ vanishes at the endpoints while the second time-derivative is nonzero. We then have $M = 1$ which suggests that the diabatic error goes as $\tau^{-4}$ for long times. 

\subsection{Diabatic error of multiple piano keys} 
\label{sec:model:subsec:multiple_pk}

We now consider the multiple piano key case as described in Sec. \ref{sec:setup:subsec:model}. Suppose that the diabatic error which results from pressing the $m$th key is denoted as $P_{m}$. In general, the total diabatic error $P$ is computed by summing over the diabatic errors corresponding to all possible transition paths which start in the ground state and end in the excited state. To simplify calculations, we restrict the number of possible transitions to be one per key. In terms of each $P_{m}$, we have
\begin{equation}
	P = \sum_{\{\vec{C}\}} \prod_{m = 1}^{n} P_{m}^{C_{m}} (1 - P_{m})^{1 - C_{m}},
	\label{eq:transProb_mpk_total}
\end{equation}
where $\vec{C}$ is a vector of length $n$ with elements $C_{m} \in \{0, 1\}$. Here, $C_{m} = 1$ corresponds to a transition while $C_{m} = 0$ otherwise. The number of transitions is always odd and leads to a restriction on the set $\{\vec{C}\}$, namely that $\sum_{m=1}^{n} C_{m} ~\text{mod}~ 2 = 1$. We note that the expression in Eq. (\ref{eq:transProb_mpk_total}) does not account for interference between different transition paths which manifest as St\"{u}ckelberg phases in the diabatic error \cite{Shevchenko2010, Ivakhnenko2023}. As such, one may view this expression as the result of averaging over the oscillations emerging from these phases. 

We assume that $P_{m}$ for each key, depending on the tuning function used, can take the form of Eq. (\ref{eq:transProb_linear}) or Eq. (\ref{eq:transProb_sin}) with the replacements $R \rightarrow R/n$ and $\tau \rightarrow \tau/n$. We emphasize that this assumption leads to $P_{m}$ being identical for each key which is not necessarily true in our simulations of the Kitaev chain; this will be discussed later. From the expected analytical form of $P_{m}$, one can see that $P_{m} \ll 1$ for sufficiently large $\tau/n$ which results in the suppression of the higher order terms $P_{m}P_{m+1}P_{m+2} ...$ in Eq. (\ref{eq:transProb_mpk_total}). This leads to a simple approximation for the total diabatic error:
\begin{equation}
	P \approx \sum_{m = 1}^{n} P_{m}.
	\label{eq:transProb_mpk_total_expand}
\end{equation}

We note that the assumption of having each key share the same diabatic error, although simple and intuitive, is generally untrue. The reason for this stems from the actual minimum gap found for each key. As shown in Fig. \ref{fig:energies_b}, the first and intermediate keys ($m = 0, 1, 2, ... n-2$) have roughly the same minimum gap which differs from the prediction of Eq. (\ref{eq:shorthand_linear}) with the replacement $R \rightarrow R/n$. Conversely, there is agreement between the minimum gap featured for the final key ($m = n-1$) and the predicted value. 

The different minimum gaps may be understood by considering the behavior of the eigenmode corresponding to $d_{1}$ as each key is pressed. We previously noted that when a key undergoes criticality, this eigenmode becomes localized within the key. The length scale over which this localization occurs is given by $R/n + \zeta$ for some $\zeta > 0$ which describes the decay of the eigenmode into neighboring sections of the chain. The final key is distinct in that it features the termination of the chain to the right. It follows that the localization length of the eigenmode in this instance is expected to be smaller compared to those of the other keys. In view of Eq. (\ref{eq:shorthand_linear}), the minimum gap may be estimated by making the replacement $R \rightarrow R/n + \zeta$. Accordingly, the minimum gap for the final key should be comparatively larger which is consistent with results from numerical calculations, see Fig. \ref{fig:energies_b}. Though it may be instructive to calculate $\zeta$ for each key, we do not discuss this in detail as our general conclusions remain unchanged provided that $\zeta$ itself is not too large.  

\section{Numerical results} \label{sec:results}

\begin{figure*}[t]
    \centering
  	\includegraphics[width=1.0\linewidth]{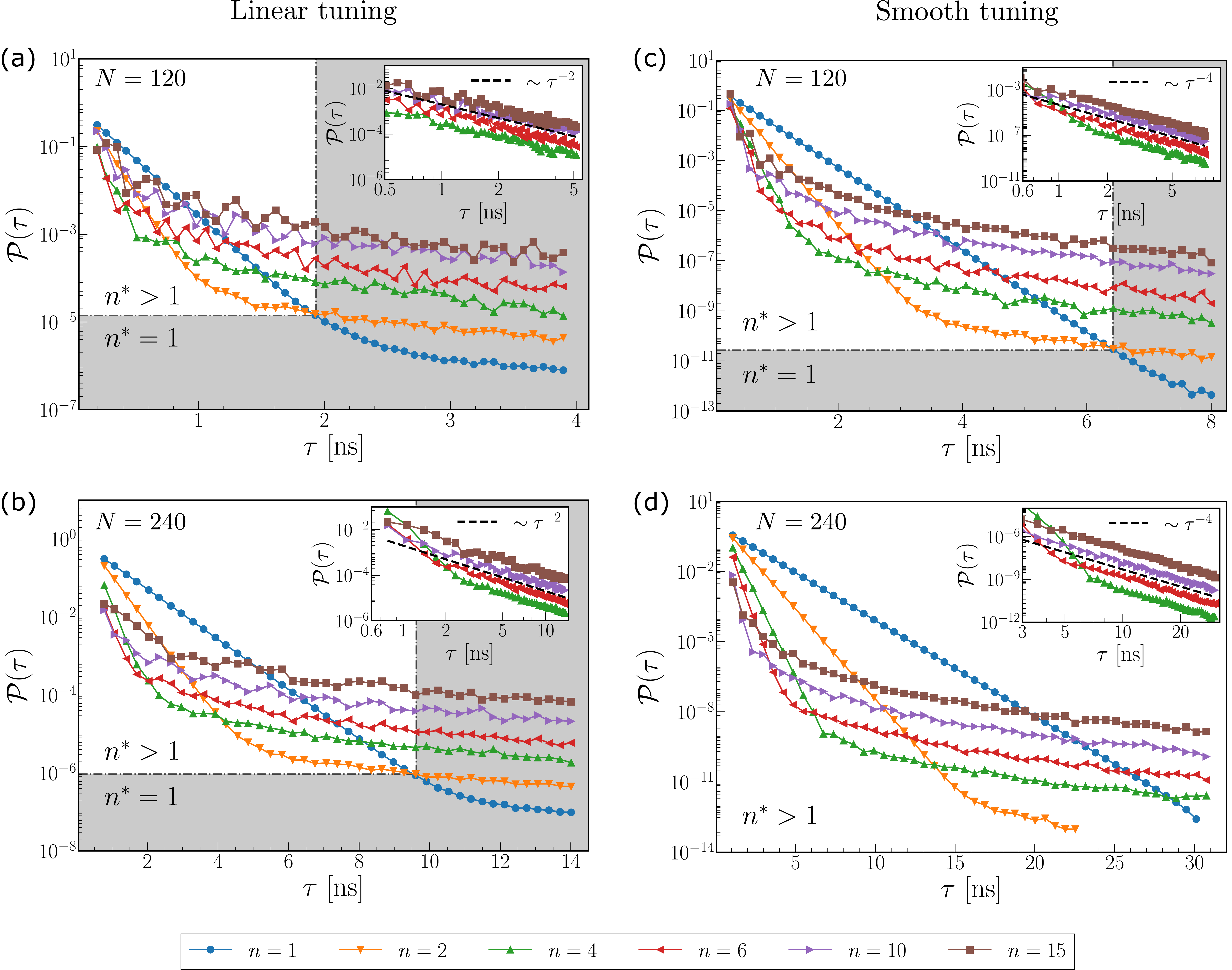}
    \caption[]{Numerical results for the diabatic error $\mathcal{P}$ as a function of the total time $\tau$ it takes to transport the MZM $\gamma_{2}$ over a fixed distance using $n$ equally sized piano keys. The results featured in the left column correspond to a linear tuning of the chemical potential in each key while the results in the right column correspond to smooth tuning. For each tuning, results are shown for a Kitaev chain with $N = 120, 240$ sites. Each plot displays an inset on log-log scale which illustrates the power-law behavior of the diabatic errors. When linear tuning is used, the diabatic error goes as $\tau^{-2}$ at large $\tau$ while for smooth tuning it goes as $\tau^{-4}$. The regions which are shaded in gray correspond to the cases where a single key is optimal for transport ($n^{*} = 1$) while within the unshaded regions, multiple keys are optimal ($n^{*} > 1$). In all simulations, we use parameters $w = 3$ meV, $\Delta_{\text{SC}} = 0.6$ meV, and $\delta \mu = 0.2$ meV.}
  	\label{fig:mainresults}
\end{figure*}

We numerically simulate the piano key transport of the MZM $\gamma_{2}$ on a Kitaev chain and calculate the diabatic error that is accumulated throughout this protocol. The diabatic error is calculated using Eq. (\ref{eq:diaberr_gen}) where the time evolution operator $U(\tau)$ is constructed using a time discretization scheme while the square of the matrix element is evaluated using covariance matrices, see Appendices \ref{app:time} and \ref{app:covariance} for details. In all simulations, the chemical potential in each key is tuned symmetrically around the critical value $w$, as detailed in Sec. \ref{sec:model:subsec:single_pk}, with $\delta \mu = 0.2$ meV. The coupling amplitudes are chosen to be $w = 3$ meV and $\Delta_{\text{SC}} = 0.6$ meV. We consider chain sizes of $N = 120$ sites and $N = 240$ sites which correspond to physical lengths $l \sim 1 ~\mu$m and $l \sim 2 ~\mu$m, respectively, for nanowires with lattice spacing $a \sim 10^{-2} ~\mu$m. The left and right chain sections are set to be equal, $L = R = N/2$. 

\begin{figure}[t]
    \centering
  	\includegraphics[width=1.0\linewidth]{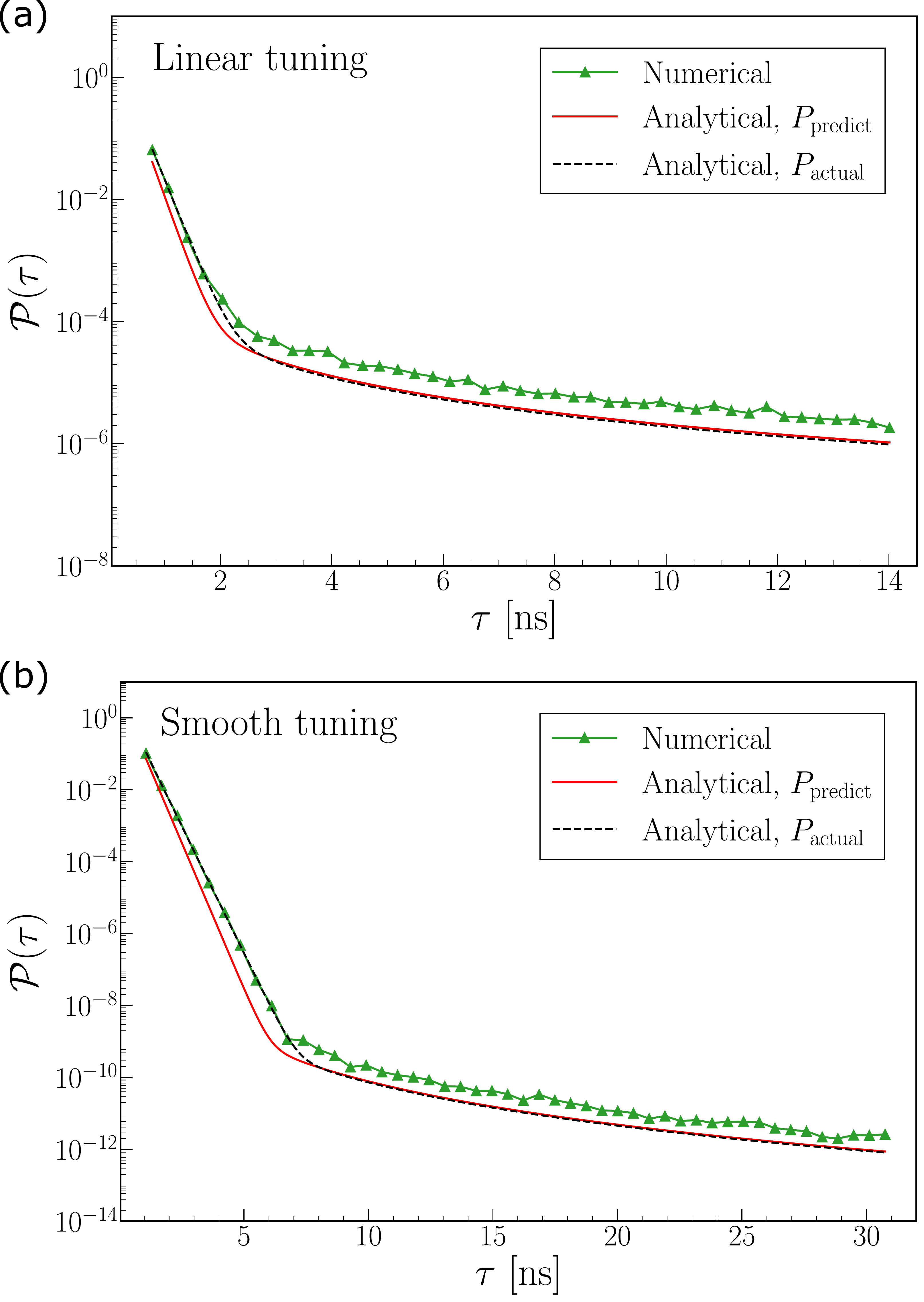}
    \caption[]{Comparison between the numerical results for the diabatic error in the case of $n = 4$ with two versions of Eq. (\ref{eq:transProb_mpk_total_expand}). The expression $P_{\text{predict}}$ (solid red curves) assumes that the minimum gap of each key is equal and given by the two-level model prediction, see Eq. (\ref{eq:shorthand_linear}). $P_{\text{actual}}$ (dashed black curves) uses the actual minimum gaps extracted from the instantaneous spectrum which we calculate numerically. The exponential decay of the diabatic error is more accurately captured by $P_{\text{actual}}$. For either expression, the power-law behavior is well-described despite being slightly underestimated. These results and expressions correspond to a chain with $N = 240$ sites. The remaining parameters used are identical to those found in the main results, see Fig. \ref{fig:mainresults}.}
  	\label{fig:results_comparison}
\end{figure}

Our main results for the diabatic error are shown in Fig. \ref{fig:mainresults} for both linear and smooth tuning functions. For each tuning, results corresponding to two different chain sizes are shown. When the chain size increases while the number of keys used $n$ is fixed, the size of each key naturally increases. We comment that each key must then be pressed for a longer amount of time to achieve a particular value for the diabatic error. This is reflected in the different time scales between results. Furthermore, we observe that regardless of the choice of tuning or choice of $n$, the behavior of the diabatic error may be divided into two distinct regimes: an exponential regime for short total times and a power-law regime for long total times. This is consistent with the analytical predictions discussed in preceding sections. We note that as $n$ increases, the power-law behavior sets in at shorter total times while the size of the exponential regime shrinks.

The tuning functions have the effect of modifying the details of the decay rate in the exponential regime as well as the precise $\tau$-dependence in the power-law regime. In the exponential regime, the diabatic error decays faster when linear tuning is used. This makes the choice of linear tuning appealing if relatively large diabatic errors are tolerated. However, for the power-law regime, linear tuning results in a $\sim \tau^{-2}$ behavior while for smooth tuning this is $\sim \tau^{-4}$. This in turn makes smooth tuning appealing if relatively small diabatic errors are desired. We remark that an ideal tuning function is one which passes slowly through the critical point and contains a large number of vanishing time-derivatives at its endpoints, starting with the first time-derivative. The former requirement results in a faster initial exponential decay while the latter leads to a larger negative power-law behavior at long total times. 

\begin{figure}[t]
    \centering
  	\includegraphics[width=1.0\linewidth]{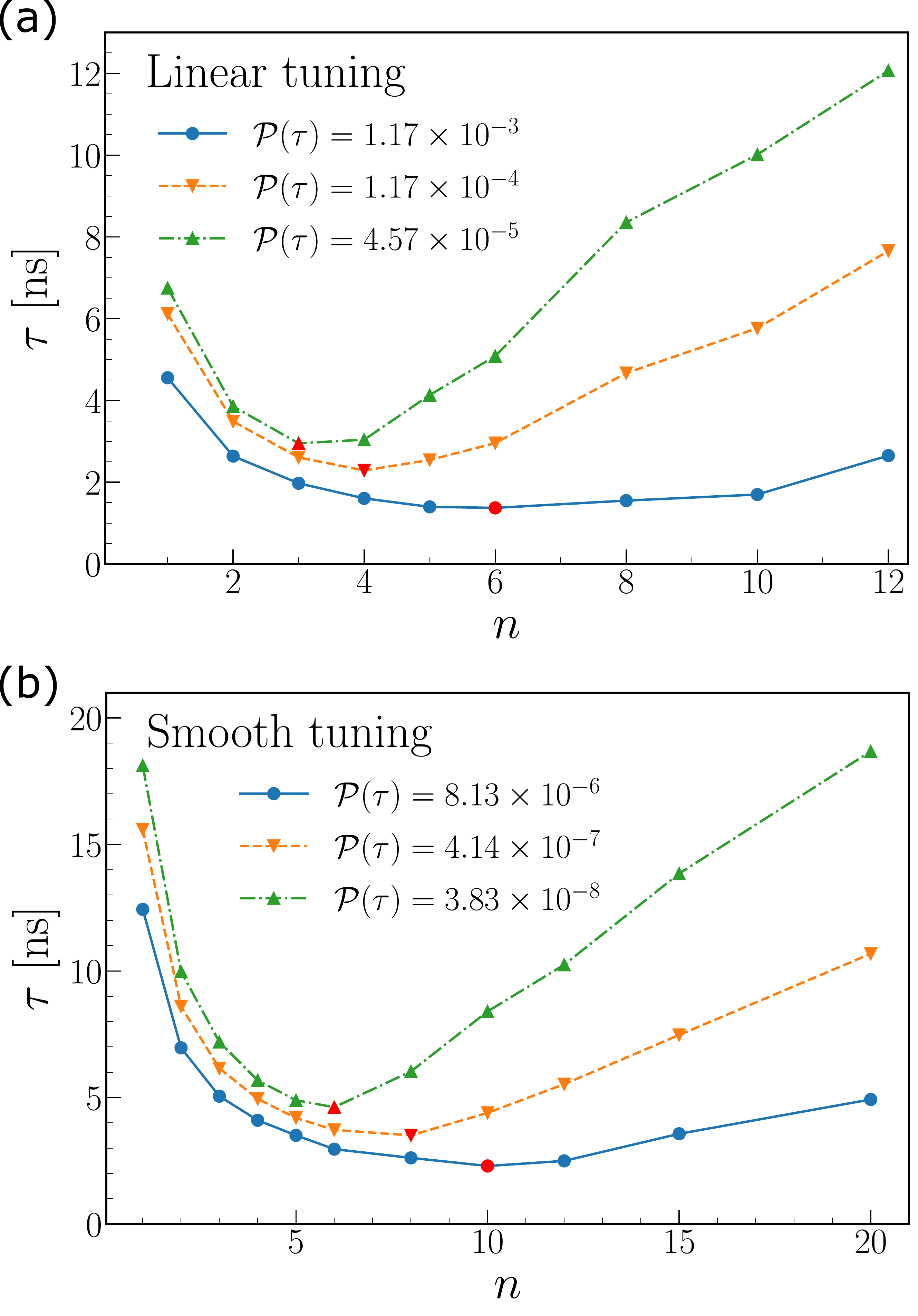}
    \caption[]{Total time as a function of the number of keys used in transport for select tolerance values of the diabatic error. The optimal values $n^{*}$ for the number of keys are highlighted as red symbols. As the tolerance for the diabatic error increases, $n^{*}$ shifts to larger values. The parameters used are identical to those found in the main results for a chain with $N = 240$ sites.}
  	\label{fig:results_tauVn}
\end{figure} 

\begin{figure}[t]
    \centering
  	\includegraphics[width=1.0\linewidth]{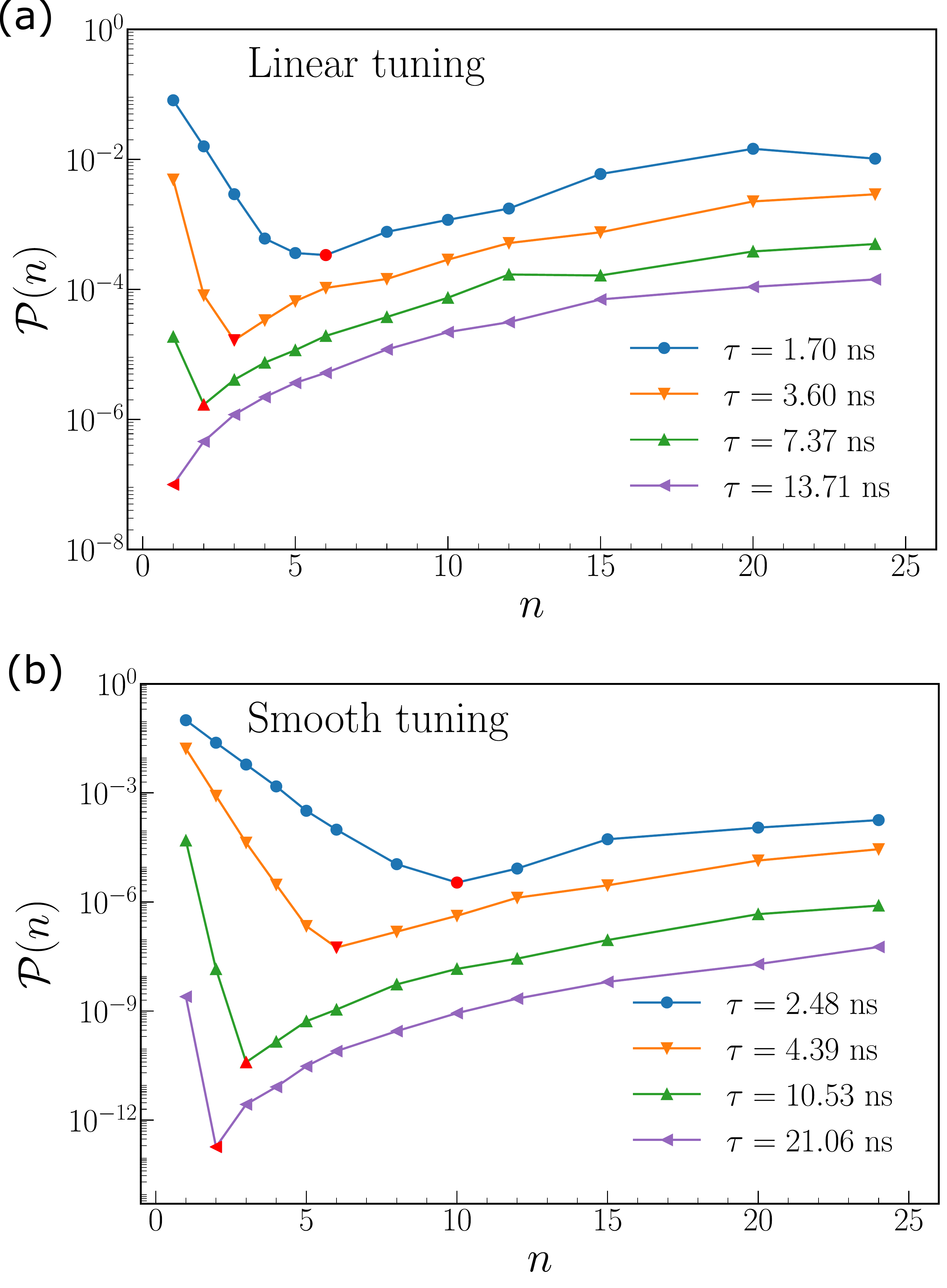}
    \caption[]{Diabatic error as a function of the number of keys used in transport for select values of the total time. The optimal values $n^{*}$ for the number of keys are highlighted as red symbols. As the total time increases, $n^{*}$ shifts to smaller values and eventually $n^{*} = 1$ is achieved when the total time is sufficiently large. The parameters used are identical to those found in the main results for a chain with $N = 240$ sites.}
  	\label{fig:results_diabVn}
\end{figure}

We compare the results for the diabatic error to the analytical expressions found in Secs. \ref{sec:model:subsec:single_pk} and \ref{sec:model:subsec:multiple_pk}. As shown in Fig. \ref{fig:results_comparison}, we examine the results corresponding to $n = 4$ as a representative example and note that similar conclusions may be drawn for all other results. Specifically, the numerical results are compared to the total diabatic error expression in Eq. (\ref{eq:transProb_mpk_total_expand}) in tandem with Eqs. (\ref{eq:transProb_linear}) and (\ref{eq:transProb_sin}) depending on the tuning function used. Two versions of Eq. (\ref{eq:transProb_mpk_total_expand}) are considered. In one version, which we denote as $P_{\text{predict}}$, the minimum gaps corresponding to each key are naively assumed to be identical and given by Eq. (\ref{eq:shorthand_linear}). In the other version, which is denoted as $P_{\text{actual}}$, the minimum gaps of each key are evaluated numerically from the instantaneous spectrum and are inserted into Eq. (\ref{eq:transProb_mpk_total_expand}). From Fig. \ref{fig:results_comparison}, we see that $P_{\text{predict}}$ fails to accurately capture the true exponential decay of the diabatic error. This is expected since, as was previously discussed in Sec. \ref{sec:model:subsec:multiple_pk}, the first and intermediate keys feature minimum gaps which are smaller compared to the predicted values. When the actual minimum gaps are used instead, we observe good agreement between $P_{\text{actual}}$ and numerical results. For either expression, the power-law behavior is described well, though an underestimate of the results is observed.

It is practical to establish a tolerance for the diabatic error and examine the number of keys that should be used such that this tolerance is achieved in the shortest amount of time. This is the motivation behind Fig. \ref{fig:results_tauVn}, which shows the total time as a function of $n$ for a selection of error tolerances. Each of the results in Fig. \ref{fig:results_tauVn} feature a minimum corresponding to a number of keys $n^{*}$. Generally, we observe that this optimal number of keys grows with the value of the error tolerance. For vanishingly small values of the tolerance, a single key becomes optimal and so $n^{*} = 1$. 

Another practical approach reverses the previous analysis and involves fixing the total time and examining the number of keys that should be used such that the diabatic error is minimized. This is demonstrated in Fig. \ref{fig:results_diabVn} where we plot the diabatic error as a function of $n$ for select values of the total time. Similar to the previous approach, an optimal number of keys $n^{*}$ emerges. As the total time increases, this optimal number falls and eventually achieves a value $n^{*} = 1$ when relatively large total times are reached.  

In view of the preceding discussion on the optimal key number $n^{*}$, whether it be with respect to the total time or diabatic error tolerance, we return to the main results shown in Fig. \ref{fig:mainresults} and categorize our results based on the value of $n^{*}$. Specifically, we divide the results into two regions based on whether $n^{*} = 1$, which represents a single key advantage (illustrated in gray in Fig. \ref{fig:mainresults}), or $n^{*} > 1$, which represents a multiple key advantage (illustrated in white). The boundary between both regions is defined using the intersection of the $n = 1$ and $n = 2$ results.

\section{Conclusion} \label{sec:conclusion}
We have studied the diabatic error that is produced as an MZM is transported across a topological superconducting wire using the piano key approach where sections of the wire are electrically gated. We specifically focused on transport which is conducted using a series of equally sized keys. Our work serves as a generalization of Ref.~\cite{Bauer2018} which considered single key transport and established that the diabatic error in this case can be adequately modelled by the Landau-Zener transition probability. In the case of multiple key transport, we demonstrated using numerical simulations that the diabatic error displays a prominent power-law behavior alongside the usual exponential decay which corresponds to the familiar Landau-Zener result. We reviewed the calculation of the diabatic error and included additional correction terms which accurately describe this power-law behavior observed. The extent of these power-law corrections depends crucially on the analyticity of the chemical potential tuning function at the beginning and end of each key press. Crucially, while the usual Landau-Zener result for the diabatic error would suggest that transport is optimal when several small (in length) keys are used instead of one large key, the power-law corrections imply the opposite. This results in a nontrivial optimal number of keys $n^{*}$ given total time or error constraints on the transport of the MZMs.  

The presence of these power-law corrections suggests that one cannot continue to exponentially suppress unwanted transitions by simply stretching the tuning function over longer times. When this is done, the power-law corrections become significant compared to the exponentially small contribution to the diabatic error. Ideally, the reduction of the diabatic error beyond the exponential regime should be done by choosing a tuning function such that its first nonvanishing time-derivative at the endpoints is of a high rank. This rank directly affects the specific power-law correction that is realized. However, we also note that in the presence of noise in the chemical potential, this may not be effective---there is an expectation that with the addition of noise, the first time-derivatives of the tuning functions at the endpoints will become discontinuous leading to a universal $\tau^{-2}$ behavior in the diabatic error as discussed in Ref.~\cite{Knapp2016}. This requires further investigation. Also, as noted above, our work considers a simplified model of transport where each key shares the same size and tuning time. Future work may consider optimizing the protocol further by lifting some of these restrictions. An examination of the ideas presented here is of great interest especially from the purview of future experiments.



\begin{acknowledgments}
The authors acknowledge financial support from the Natural Sciences and Engineering Research Council of Canada (NSERC) and the Fonds de Recherche du Qu\'{e}bec --- Nature et technologies. BPT acknowledges support through the Postgraduate Scholarships --- Doctoral program of NSERC.
\end{acknowledgments}

\appendix
\section{Kitaev chain in the Majorana basis} \label{app:majorana}
We review certain details of the Kitaev chain which are relevant in the numerical calculation of the diabatic error. First, it is convenient to recast the Hamiltonian for the Kitaev chain in Eq. (\ref{eq:ham_kitaev_realspace}) in terms of Majorana operators:
 \begin{align}
	H = &-\frac{i}{2} \sum_{j = 1}^{N} \mu_{j} \gamma_{j}^{A} \gamma_{j}^{B} + \frac{i}{4} (\Delta_{\text{SC}} - w) \sum_{j=1}^{N-1} \gamma_{j}^{A} \gamma_{j+1}^{B} 
	\label{eq:ham_kitaev_realspace_maj}
	\\
	&+ \frac{i}{4}(\Delta_{\text{SC}} + w) \sum_{j = 1}^{N-1} \gamma_{j}^{B} \gamma_{j+1}^{A},
	\nonumber
\end{align}
where $\gamma_{j}^{A} = c_{j} + c_{j}^{\dagger}$ and $\gamma_{j}^{B} = -i(c_{j} - c_{j}^{\dagger})$. In the Bogoliubov de-Gennes (BdG) form with the Nambu vector $\vec{\Psi} = (\gamma_{1}^{A}, \gamma_{1}^{B}, \gamma_{2}^{A}, \gamma_{2}^{B}, ..., \gamma_{N}^{A}, \gamma_{N}^{B})^{\text{T}}$, Eq. (\ref{eq:ham_kitaev_realspace_maj}) reads
\begin{equation}
	H = \frac{i}{2} \vec{\Psi}^{\dagger} \mathcal{A} \vec{\Psi},
	\label{eq:ham_kitaev_maj_bdg}
\end{equation}
where 
\begin{equation}
	\mathcal{A} = 
	\begin{pmatrix}
		A_{1} & B & & & &
		\\
		-B & A_{2} & B & & &
		\\
		& -B & A_{3} & B & &
		\\
		& & \ddots & \ddots & \ddots
		\\
		& & & -B & A_{N-1} & B
		\\
		& & & & -B & A_{N}
	\end{pmatrix},
	\label{eq:matA}
\end{equation}
with $A_{j} = -i(\mu_{j}/2) \sigma^{y}$ and $B = (\Delta_{\text{SC}}/4)\sigma^{x} - i(w/4) \sigma^{y}$. Here, $\sigma^{i}, i = x,y,z$ are Pauli matrices. Since the matrix $\mathcal{A}$ is real and skew-symmetric, there exists a real, orthogonal transformation $\mathcal{W}$ such that $\mathcal{A}$ is brought into a block-diagonal form \cite{Kitaev2001}. Namely, 
\begin{equation}
	\mathcal{A} = \mathcal{W}^{\text{T}} \left( \bigoplus_{j = 1}^{N} \varepsilon_{j} i \sigma^{y} \right) \mathcal{W},
	\label{eq:A_blockdiag}
\end{equation}
where $\pm i \varepsilon_{j}$, $\varepsilon_{j} > 0$ are the eigenvalues of $\mathcal{A}$. In our numerics, we determine the transformation $\mathcal{W}$ using the methods of Ref. \cite{Wimmer2012}. 

By inserting Eq. (\ref{eq:A_blockdiag}) into the Hamiltonian in Eq. (\ref{eq:ham_kitaev_maj_bdg}), a transformation of the Nambu vector $\vec{\Psi}$ can be identified as $\vec{\Phi} = \mathcal{W} \vec{\Psi}$. The vector $\vec{\Phi}$ contains operators which are Majorana in character since they obey the same anticommutation relations. Denoting these new operators as $\eta_{k}$, the Hamiltonian can be expressed in a canonical form:
\begin{equation}
	H = -\frac{1}{2}\vec{\Phi}^{\dagger} \left( \bigoplus_{j = 1}^{N} \varepsilon_{j} \sigma^{y} \right)  \vec{\Phi} = i \sum_{j = 1}^{N} \varepsilon_{j} \eta_{2j-1} \eta_{2j}.
	\label{eq:ham_kit_canonical}
\end{equation}
We note that the usual Bogoliubov operators can be defined from these Majorana operators, namely $d_{j} = (1/2)(\eta_{2j-1} + i \eta_{2j})$. These operators along with the orthogonal transformation $\mathcal{W}$ will become important when we discuss the covariance matrix formalism.

\section{Time evolution operator} \label{app:time}
We describe the method by which the time evolution operator is constructed in our numerical simulations. In general, the time evolution operator takes the form 
\begin{equation}
	U(t) = \mathcal{T} \exp \left( -\frac{i}{\hbar} \int_{0}^{t} dt^{\prime} H(t^{\prime}) \right),
	\label{eq:timeevol_gen}
\end{equation}
where $\mathcal{T}$ is the time ordering operator. The single-particle counterpart of Eq. (\ref{eq:timeevol_gen}), which we denote as $\mathcal{U}(t)$, is what we compute in simulations. To determine $\mathcal{U}(t)$, let us consider the evolution of a general Majorana operator:
\begin{equation}
	\gamma = \vec{v} \cdot \vec{\Psi} = \sum_{j = 1}^{N}(v_{2j-1} \gamma_{j}^{A} + v_{2j} \gamma_{j}^{B}),
	\label{eq:gamma_gen}
\end{equation}
where $\vec{v} = (v_{1}, v_{2}, \dots, v_{2N})^{\text{T}}$ contains complex coefficients. The time evolution of $\gamma$ is given by 
\begin{equation}
	\gamma \rightarrow U(t)^{\dagger} \gamma U(t).
	\label{eq:gamma_gen_timeevol}
\end{equation}
To motivate the specific form of $\mathcal{U}(t)$, we consider the simple case where the Hamiltonian $H$ is time-independent. Then, Eq. (\ref{eq:gamma_gen_timeevol}) may be expanded using the Baker-Campbell-Hausdorff formula:
\begin{flalign}
	&U(t)^{\dagger} \gamma U(t) = \exp \left( \frac{i}{\hbar} H t \right) \gamma \exp \left( -\frac{i}{\hbar} H t \right), &&
	\label{eq:gamma_gen_timeevol2}
	\\
	&= \gamma + \left[-\frac{i}{\hbar} H t, \gamma \right] + \frac{1}{2!} \left[-\frac{i}{\hbar} H t, \left[-\frac{i}{\hbar} H t, \gamma\right]\right] + \cdots. &&
	\nonumber
\end{flalign}
By invoking the form of the Hamiltonian in Eq. (\ref{eq:ham_kitaev_maj_bdg}) along with the Majorana anticommutation relations $\{\gamma_{j}^{J}, \gamma_{j^{\prime}}^{J^{\prime}} \} = 2 \delta_{j j^{\prime}} \delta_{J J^{\prime}}$, one can show that the first commutator in the expansion becomes
\begin{equation}
	[-i H t, \gamma] = 2 t (\mathcal{A} \vec{v}) \cdot \vec{\Psi}.
	\label{eq:commute_Hgamma}
\end{equation}
Through iteration, the higher-order commutators are similarly evaluated and the following result is deduced:
\begin{equation}
	U(t)^{\dagger} \gamma U(t) = e^{2 \mathcal{A} t / \hbar} \vec{v} \cdot \vec{\Psi}.
	\label{eq:gamma_gen_timeevol3}
\end{equation}
The matrix exponential $e^{2 \mathcal{A} t / \hbar}$ plays the role of a single-particle time evolution operator since it acts directly on the coefficients $\vec{v}$ which characterize the Majorana operator. Generalizing this result to a time-dependent $H$, we have that 
\begin{equation}
	\mathcal{U}(\tau) = \mathcal{T} \exp \left( \frac{2}{\hbar} \int_{0}^{\tau} dt \mathcal{A}(t) \right),
	\label{eq:timeevol_sp_gen}
\end{equation}
where the total transport time $\tau$ has been reintroduced. In simulations, $\mathcal{U}(\tau)$ is approximated by discretizing time and taking a time ordered product of individual matrix exponentials. This is given by
\begin{equation}
	\mathcal{U}(\tau) \approx \mathcal{T} \prod_{p = 1}^{N_{s}} \exp \left( \frac{2}{\hbar} \Delta t \mathcal{A}(t_{p})  \right),
	\label{eq:timeevol_sp_discrete}
\end{equation}
where $\Delta t$ is the size of a time step and $N_{s}$ is the number of time steps. We use $\Delta t \sim 10^{-5}$ ns and determine the number of steps $N_{s}$ accordingly for each simulation such that $\tau = N_{s} \Delta t$. Typical values for the number of time steps range from $N_{s} \sim 10^{4} - 10^{5}$.

\section{Covariance matrix formalism} \label{app:covariance}
We demonstrate the application of covariance matrices in the calculation of the diabatic error. This section uses the results of Ref. \cite{Bravyi2017} and we remark that additional details regarding covariance matrices are detailed therein. Let us consider the ground state $| \Omega \rangle$ of the Kitaev chain. The elements of the covariance matrix corresponding to the ground state are given by 
\begin{equation}
	\mathcal{M}_{pq} = -\frac{i}{2} \langle \Omega | [\eta_{p}, \eta_{q}] | \Omega \rangle,
	\label{eq:covariance_GS}
\end{equation}
where $\eta_{p}$ are the Majorana operators found in the canonical form of the Hamiltonian in Eq. (\ref{eq:ham_kit_canonical}). By definition, the covariance matrix is real, contains only zeros along its diagonal, $\mathcal{M}_{pp} = 0$, and is skew-symmetric, $\mathcal{M}_{pq} = -\mathcal{M}_{qp}$. 

By using the Bogoliubov operators $d_{k}$ introduced in Appendix \ref{app:majorana}, one can show that the covariance matrix in the canonical basis of operators $\vec{\Phi}$ takes the simple form
\begin{equation}
	\mathcal{M} = \bigoplus_{j = 1}^{N} i \sigma^{y}.
	\label{eq:covariance_GS_evaled}
\end{equation}
It is useful to consider a change of basis. In particular, we switch to the basis of the original Majorana operators $\vec{\Psi}$ using the orthogonal transformation $\mathcal{W}$ from Appendix \ref{app:majorana}. In view of the definition in Eq. (\ref{eq:covariance_GS}), we have
\begin{equation}
	\mathcal{M}_{pq} = \sum_{p' q'} \mathcal{W}_{pp'} \mathcal{W}_{qq'} \left(-\frac{i}{2} \langle \Omega | [\gamma_{p^{\prime}}, \gamma_{q^{\prime}}] | \Omega \rangle \right).
	\label{eq:covariance_GS_basischange}
\end{equation}
Defining $\mathcal{M}_{0, pq} = -(i/2) \langle \Omega | [\gamma_{p}, \gamma_{q}] | \Omega \rangle$, the following matrix equation is obtained:
\begin{equation}
	\mathcal{M} = \mathcal{W} \mathcal{M}_{0} \mathcal{W}^{T}.
	\label{eq:covariance_GS_basischange_MAT}
\end{equation}
The time evolution of the covariance matrix $\mathcal{M}_{0}$ in this original Majorana basis is straightforward:
\begin{equation}
	\mathcal{M}_{0}(t) = \mathcal{U}(t)^{\dagger} \mathcal{M}_{0} \mathcal{U}(t),
	\label{eq:covariance_GS_ogbasis_timeevol}
\end{equation}
where $\mathcal{U}(t)$ is a single-particle time evolution operator.

Recall from Eq. (\ref{eq:diaberr_gen}) that the diabatic error requires the evaluation of $|\langle \Omega_{f} | U(\tau) | \Omega_{i} \rangle |^2$. This quantity can be viewed as the squared overlap between the final ground state $| \Omega_{f} \rangle$ and the time-evolved initial ground state $U(\tau) | \Omega_{i} \rangle$. In terms of covariance matrices, this overlap is given by
\begin{equation}
	|\langle \Omega_{f} | U(\tau) | \Omega_{i} \rangle |^2 = |2^{-N} \text{Pf}(\mathcal{M}_{0,f} + \mathcal{M}_{0,i}(\tau))|,
	\label{eq:covariance_diabaticerr}
\end{equation}
where $\mathcal{M}_{0,i}(t)$ and $\mathcal{M}_{0,f}$ are the covariance matrices of the time-evolved initial ground state and final ground state, respectively, while $\text{Pf}( \cdots )$ denotes the Pfaffian. Note that $\mathcal{M}_{0,i}(t)$ is determined using the covariance matrix of the initial ground state, $\mathcal{M}_{0,i}$, in conjunction with Eq. (\ref{eq:covariance_GS_ogbasis_timeevol}).

\end{document}